\documentclass[graybox]{svmult}

\usepackage{mathptmx}
\usepackage{helvet}
\usepackage{courier}
\usepackage{type1cm}
\usepackage{graphicx}
\usepackage{amsmath,amssymb}
\usepackage{booktabs}
\usepackage{tabularx}
\usepackage{url}
\usepackage[bottom]{footmisc}

\graphicspath{{figures/}}
\begin{document}

\title*{Learning-Based Power Flow Estimation with Sparse PMU Deployment in Power Systems}
\titlerunning{Learning-Based Power Flow Estimation with Sparse PMU Deployment}

\author{Van-Hai Bui, Quang Manh Hoang, Akhtar Hussain, and Wencong Su}
\authorrunning{Van-Hai Bui et al.}

\institute{Van-Hai Bui, Quang Manh Hoang, and Wencong Su \at
Department of Electrical and Computer Engineering, University of Michigan-Dearborn, Dearborn, Michigan, USA,\\
\email{vhbui@umich.edu, manhhq@umich.edu, wencong@umich.edu}
\and
Akhtar Hussain \at
Department of Electrical and Computer Engineering, Laval University, Quebec City, Canada,\\
\email{akhtar.hussain@gel.ulaval.ca}
\and
Corresponding author: Van-Hai Bui, \email{vhbui@umich.edu}}

\maketitle

\abstract{Accurate monitoring of power flow is fundamental to the secure and reliable operation of modern power systems. Existing data-driven power flow estimation methods typically assume that synchronized measurements are available at every bus, requiring extensive deployment of phasor measurement units (PMUs). Such dense sensing infrastructures substantially increase installation, communication, and maintenance costs. This paper proposes a learning-based framework that minimizes the number of installed PMUs while maintaining high-accuracy system-wide power flow estimation. Instead of relying on measurements from all buses, the proposed method identifies an informative subset of buses whose synchronized measurements are sufficient to accurately reconstruct branch power flow across the network. To this end, a neural network is developed to learn the nonlinear relationship between PMU measurements and system-wide power flows, and multiple PMU deployment scenarios are evaluated to quantify the trade-off between measurement availability and estimation accuracy. The results enable utilities to identify cost-effective PMU placement strategies that balance deployment cost and prediction performance. Case studies on the IEEE 30-bus test system demonstrate that the proposed framework significantly reduces the number of required PMUs while maintaining accurate power flow estimation.}


\section{Introduction}

Accurate knowledge of bus voltages, power injections, and branch power flows is fundamental to the secure, reliable, and economical operation of modern power systems. Phasor measurement units (PMUs) provide time-synchronized measurements that enable operators to observe voltage magnitudes, phase angles, frequencies, and current phasors across the network using a common time reference \cite{deLaRee2010,phadke2018,aminifar2015}. These synchronized measurements have become an essential data source for state estimation, disturbance detection, stability assessment, fault location, oscillation monitoring, protection, and wide-area control \cite{chen2006,cheng2024}. PMU measurements can also be integrated into conventional state estimation to improve estimation accuracy and numerical robustness \cite{zhou2006}. However, installing PMUs at every bus is generally impractical because each installation requires measurement hardware, communication infrastructure, data storage, cybersecurity, maintenance, and integration with existing utility systems. Consequently, utilities must balance the benefits of improved system observability against the cost and complexity of deploying a dense PMU network.

The increasing availability of synchronized measurements has also accelerated the development of data-driven techniques for power-system monitoring and analysis, particularly for systems with limited PMU coverage. Machine-learning models can learn the nonlinear relationship between available measurements and unobserved system variables from historical or simulated operating data. These approaches have been applied to power flow approximation, dynamic state estimation, voltage-stability assessment, and event detection. For example, neural-network-based power flow models have been developed to approximate the mapping between load and generation conditions and the corresponding power flow solutions \cite{pham2022}. Physics-informed neural networks further improve generalization by incorporating network equations or physical constraints into the training process \cite{hu2020}. Likewise, neural networks and graph neural networks have been employed for state estimation using conventional and synchronized measurements, even under incomplete PMU coverage \cite{varghese2024,kundacina2022}. PMU data have also been combined with machine-learning techniques for forced-oscillation detection, short-term voltage-stability assessment, and the development of benchmark event datasets \cite{ayachi2020,li2023,foggo2023}. More recently, Yue et al. developed a graph-learning-assisted monitoring framework that combines sparse fast measurements from PMUs and SCADA with slow but abundant AMI measurements to construct synchronized measurement snapshots for robust distribution-system state estimation \cite{yue2024}. However, this framework focuses on measurement synchronization, system-state reconstruction, and event detection under heterogeneous data streams. It does not investigate how the number and locations of synchronized PMU measurements affect the accuracy of direct system-wide branch power-flow estimation.

Despite these advances, most existing learning-based estimation methods assume that measurements are available from a large number of buses or from a fixed sensing configuration. This assumption is often unrealistic in practical systems where PMU penetration remains limited. More importantly, estimation accuracy depends not only on the number of installed PMUs but also on their locations. Measurements collected at electrically informative buses capture network-wide operating conditions more effectively than those collected at less informative locations. Consequently, two PMU configurations with the same number of devices can produce significantly different estimation performance. Although recent studies have demonstrated that machine-learning methods can estimate power flows and power injections in PMU-unobservable systems, determining both the optimal number and the optimal locations of PMUs for learning-based estimation remains an important research challenge \cite{sahoo2023}.

The PMU placement problem has been extensively investigated from the perspectives of topological and numerical observability. Early studies formulated the problem as identifying the minimum number of PMUs required to achieve complete network observability \cite{baldwin1993}. Subsequent research incorporated practical considerations such as branch outages, measurement redundancy, zero-injection buses, bad-data detection, limited PMU channels, and installation costs. A variety of optimization techniques, including binary search, integer linear programming, convex relaxation, graph-theoretic methods, and metaheuristic algorithms, have been proposed for PMU placement \cite{kekatos2012,johnson2021}. While these studies provide a solid foundation for cost-effective PMU deployment, they primarily focus on minimizing the number of PMUs or maximizing system observability. The trade-off among the number of PMUs, their locations, and the resulting accuracy of learning-based power flow estimation has received comparatively little attention.

To address this gap, this study proposes a learning-based framework to investigate the impact of PMU deployment on system-wide branch power-flow estimation. The proposed framework evaluates how both the number and locations of monitored buses influence estimation accuracy. Active and reactive power measurements collected from a selected subset of buses are used as the inputs to a neural network, while the active power flows of all branches are estimated as the outputs. Multiple PMU deployment scenarios are systematically evaluated to identify sensing configurations that maintain high estimation accuracy while reducing the number of monitored buses. Unlike conventional PMU placement methods that primarily target complete network observability, the proposed framework directly links PMU deployment to the performance of a data-driven power flow estimator.

The main contributions of this paper are summarized as follows:
\begin{itemize}
    \item A learning-based power-flow estimation framework is developed to estimate system-wide branch power flows using active and reactive power measurements collected from only a subset of buses.
    \item The impacts of both the number and locations of PMUs on estimation accuracy are systematically investigated, providing quantitative insights into the trade-off between measurement availability and prediction performance.
    \item The proposed framework is validated on the IEEE 30-bus test system under diverse loading and generation scenarios to evaluate the robustness of the neural-network-based estimator.
    \item A comprehensive analysis of PMU deployment strategies is presented, identifying the most informative measurement locations and quantifying the degradation in estimation accuracy as PMU availability is reduced.
\end{itemize}

\section{Problem Formulation}

\subsection{Power flow model}

In this study, a power-flow model is employed to generate the training dataset and validate the proposed machine-learning framework for branch power flow estimation. The model is based on the conventional power-flow equations, which describe the relationship between bus voltages, power injections, and network admittance. The complex power injection at bus $k$ is expressed as
\begin{equation}
\begin{aligned}
S_k=P_k+jQ_k
=|V_k|\sum_{j=1}^{N}|V_j|\big[&G_{kj}\cos\theta_{kj}+B_{kj}\sin\theta_{kj}\\
&+j\left(G_{kj}\sin\theta_{kj}-B_{kj}\cos\theta_{kj}\right)\big].
\end{aligned}
\label{eq:complex_power}
\end{equation}
where $V_k$ is the voltage magnitude at bus $k$, $Y_{ik}=G_{ik}+jB_{ik}$ is the admittance between buses $i$ and $k$, $\theta_k$ is the voltage phase angle at bus $k$, $P_k$ is the active power injected at bus $k$, and $Q_k$ is the reactive power injected at bus $k$.

The active and reactive power injections at bus $k$ are obtained from \eqref{eq:complex_power} and are expressed as
\begin{equation}
P_k=|V_k|\sum_{j=1}^{N}|V_j|\left[G_{kj}\cos\left(\theta_k-\theta_j\right)+B_{kj}\sin\left(\theta_k-\theta_j\right)\right],
\label{eq:active_power}
\end{equation}
\begin{equation}
Q_k=|V_k|\sum_{j=1}^{N}|V_j|\left[G_{kj}\sin\left(\theta_k-\theta_j\right)-B_{kj}\cos\left(\theta_k-\theta_j\right)\right].
\label{eq:reactive_power}
\end{equation}

\begin{figure}[t]
\centering
\includegraphics[width=0.62\textwidth]{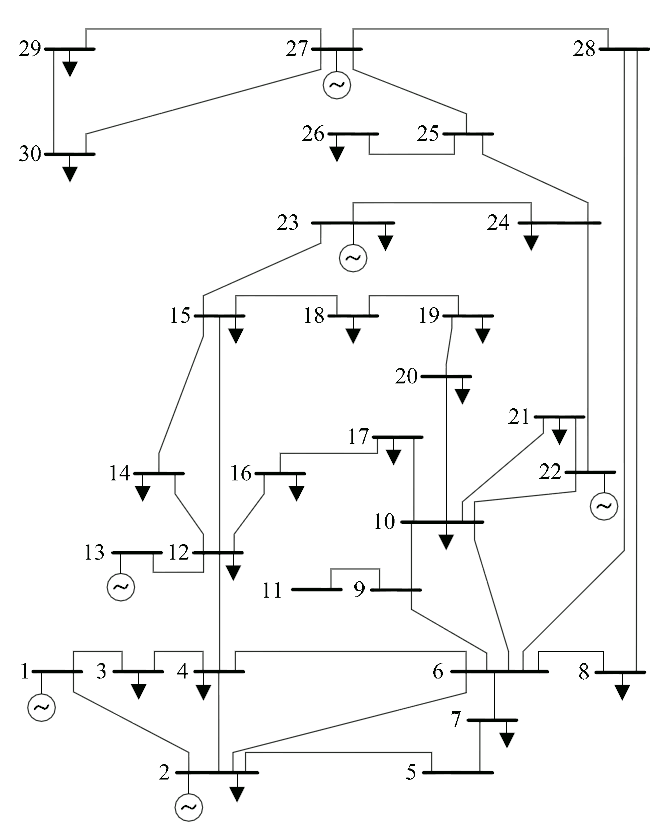}
\caption{IEEE 30-bus test system.}
\label{fig:ieee30}
\end{figure}

\begin{figure}[t]
\centering
\includegraphics[width=0.70\textwidth]{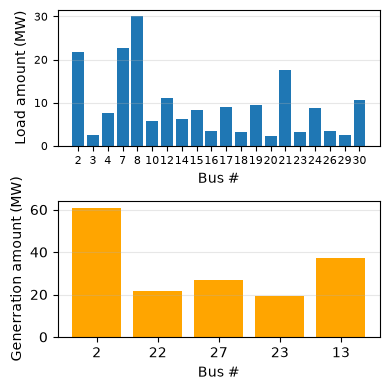}
\caption{Base-case active load and generation profiles.}
\label{fig:base_profiles}
\end{figure}

The IEEE 30-bus test system \cite{uwTestCaseArchive}, shown in Fig.~\ref{fig:ieee30}, is adopted to evaluate the proposed framework. To construct a comprehensive dataset, 50,000 operating scenarios are generated by randomly varying the active and reactive loads as well as the generator outputs. The corresponding power flow equations are solved for each operating scenario to obtain the branch power flows, which serve as the ground-truth outputs for training and testing the neural-network model. The base-case load and generation profiles are illustrated in Fig.~\ref{fig:base_profiles}.

\subsection{Dataset generation and visualization}

The dataset is generated by repeatedly solving the power flow equations in \eqref{eq:complex_power}--\eqref{eq:reactive_power} under diverse operating conditions. A total of 50,000 operating scenarios are simulated in the Python environment using the pandapower library \cite{pandapower}. In each scenario, the active and reactive loads as well as the generator outputs are randomly varied within $\pm50\%$ of their base-case values, and a power-flow analysis is performed to obtain the corresponding branch power flows. The input features consist of the active and reactive power measurements at the monitored buses, while the target outputs are the power flow of all transmission branches.

\begin{figure}[t]
\centering
\includegraphics[width=0.72\textwidth]{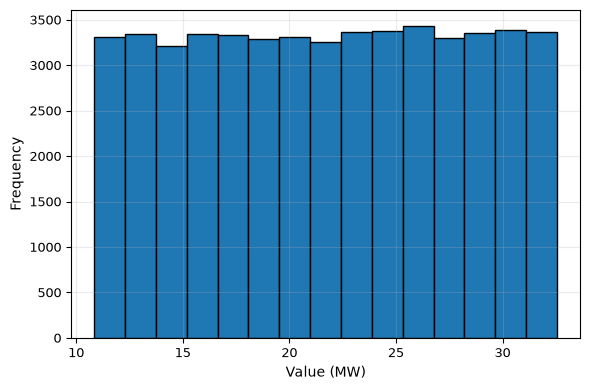}
\caption{Distribution of active power demand at bus 2.}
\label{fig:bus2_distribution}
\end{figure}

\begin{figure}[t]
\centering
\includegraphics[width=0.72\textwidth]{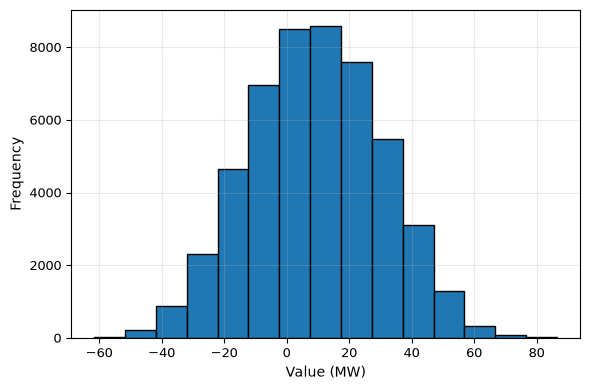}
\caption{Distribution of power flow from bus 1 to bus 2.}
\label{fig:line12_distribution}
\end{figure}

\begin{figure}[t]
\centering
\includegraphics[width=0.78\textwidth]{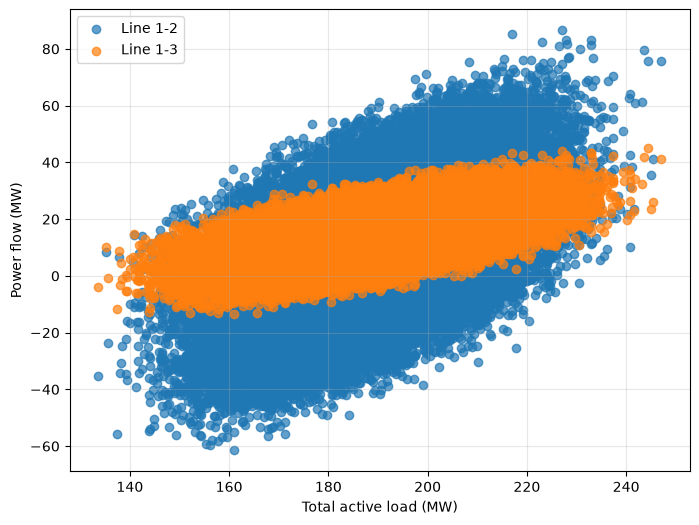}
\caption{Total load versus line power flow.}
\label{fig:load_flow_relation}
\end{figure}

The complete dataset represents the case where PMUs are installed at every bus. To investigate the impact of limited PMU deployment, different PMU configurations are created by removing the measurements from unmonitored buses before they are provided as inputs to the neural-network model. This process enables a systematic evaluation of how both the number and locations of PMUs affect power flow estimation accuracy. Figures~\ref{fig:bus2_distribution}--\ref{fig:load_flow_relation} present data distributions at selected buses, including the active power demand at bus 2, the active power flow from bus 1 to bus 2, and the relationship between the total system load and the power flow on the first transmission line, respectively.

\subsection{Learning-based power flow estimation problem}

To estimate branch power flows from PMU measurements, a fully connected neural network is developed to learn the nonlinear mapping between the system operating conditions and the corresponding branch power flows. The input vector consists of the active and reactive power measurements collected from the monitored buses. In addition, the active power outputs of generator buses are included as input features, while the slack-bus active power injection is excluded because it is determined by the power flow solution. The input vector is expressed as
\begin{equation}
\mathbf{x}=\left[P_2,\ldots,P_N,\;Q_1,Q_2,\ldots,Q_N,\;P_{G1},P_{G2},\ldots,P_{GM}\right]^{T},
\label{eq:input_vector}
\end{equation}
where $N$ is the total number of buses and $M$ is the number of generator buses.

The IEEE 30-bus test system contains 41 transmission branches. The neural network predicts the active and reactive power flows of all branches simultaneously, and the output vector is given by
\begin{equation}
\mathbf{y}_{\mathrm{branch}}=\left[P_1,\ldots,P_{41},\;Q_1,\ldots,Q_{41}\right]^{T}.
\label{eq:output_vector}
\end{equation}

Consider a fully connected neural network with $L$ layers. Let $\mathbf{h}^{(0)}=\mathbf{x}$ denote the input layer. The hidden layers are computed as
\begin{equation}
\mathbf{z}^{(l)}=\mathbf{W}^{(l)}\mathbf{h}^{(l-1)}+\mathbf{b}^{(l)},
\qquad l=1,\ldots,L-1,
\label{eq:nn_linear}
\end{equation}
\begin{equation}
\mathbf{h}^{(l)}=\sigma_l\!\left(\mathbf{z}^{(l)}\right),
\qquad l=1,\ldots,L-1,
\label{eq:nn_activation}
\end{equation}
where $\mathbf{W}^{(l)}$ is the weight matrix, $\mathbf{b}^{(l)}$ is the bias vector, and $\sigma_l(\cdot)$ is the activation function. The output layer is expressed as
\begin{equation}
\widehat{\mathbf{y}}=\mathbf{W}^{(L)}\mathbf{h}^{(L-1)}+\mathbf{b}^{(L)}.
\label{eq:nn_output}
\end{equation}

\begin{figure}[t]
\centering
\includegraphics[width=\textwidth]{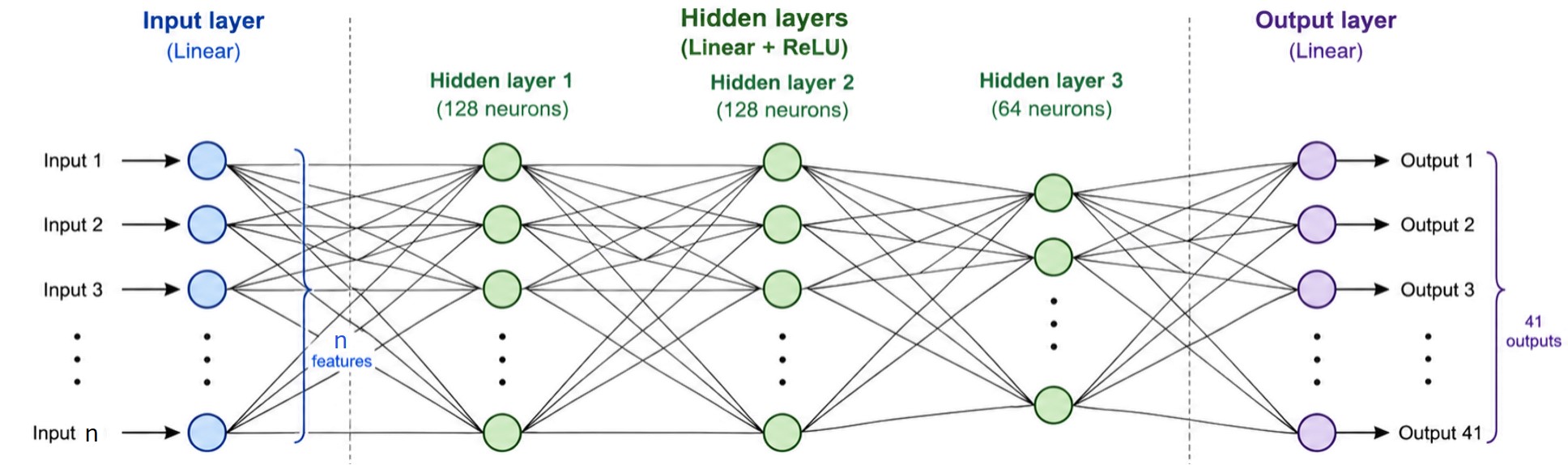}
\caption{Architecture of the proposed neural-network-based power flow estimation framework.}
\label{fig:nn_architecture}
\end{figure}

The overall architecture of the proposed neural-network-based power flow estimator is illustrated in Fig.~\ref{fig:nn_architecture}. The model takes PMU measurements, including active and reactive load injections and active generator outputs, as inputs and predicts the power flows of all transmission branches. The generated dataset is divided into 32,000 training samples, 8,000 validation samples, and 10,000 testing samples to train and evaluate the model.

\begin{itemize}
    \item Inputs: PMU measurements, including active/reactive load injections ($P/Q$) and active power outputs of generator buses.
    \item Outputs: Power flows of all transmission branches (41 outputs for the IEEE 30-bus system).
\end{itemize}

\subsection{Performance metrics}

The performance of the proposed neural network model is evaluated using three widely adopted regression metrics: mean squared error (MSE), root mean squared error (RMSE), and the coefficient of determination ($R^2$). These metrics provide complementary perspectives on the prediction accuracy of the estimated power flows.

The MSE quantifies the average squared difference between the predicted and actual branch power flows. It is defined as
\begin{equation}
\mathrm{MSE}=\frac{1}{N}\sum_{i=1}^{N}\left(y_i-\widehat{y}_i\right)^2,
\label{eq:mse}
\end{equation}
where $N$ is the total number of testing samples, $y_i$ is the actual branch power flow obtained from the power flow solver, and $\widehat{y}_i$ is the corresponding prediction generated by the model.

The RMSE is also adopted and is defined as
\begin{equation}
\mathrm{RMSE}=\sqrt{\frac{1}{N}\sum_{i=1}^{N}\left(y_i-\widehat{y}_i\right)^2}.
\label{eq:rmse}
\end{equation}
RMSE provides a direct measure of the average prediction error. Smaller RMSE values indicate higher estimation accuracy.

The coefficient of determination evaluates the proportion of variance in the actual branch power flows that is explained by the prediction model. It is defined as
\begin{equation}
R^2=1-\frac{\sum_{i=1}^{N}\left(y_i-\widehat{y}_i\right)^2}{\sum_{i=1}^{N}\left(y_i-\overline{y}\right)^2},
\label{eq:r2}
\end{equation}
where $\overline{y}=N^{-1}\sum_{i=1}^{N}y_i$ is the mean of the actual observations. The value $R^2=1$ indicates perfect agreement between the predicted and actual branch power flows, whereas $R^2=0$ indicates that the model performs no better than simply predicting the mean of the observations.

\section{Results and Discussion}

\subsection{Baseline performance using full PMU deployment}

\begin{figure}[t]
\centering
\includegraphics[width=0.72\textwidth]{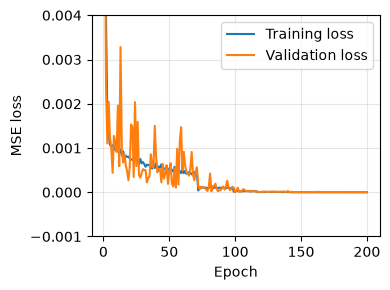}
\caption{Training and validation loss history.}
\label{fig:loss_history}
\end{figure}

To establish a performance benchmark, the proposed neural network is first evaluated using the complete set of PMU measurements from all 30 buses. The model is trained for 200 epochs using a batch size of 16 and a learning rate of 0.001. The training and validation loss curves are shown in Fig.~\ref{fig:loss_history}, demonstrating stable convergence without noticeable overfitting.

\begin{figure}[t]
\centering
\includegraphics[width=0.72\textwidth]{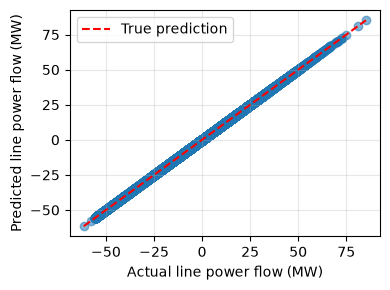}
\caption{Comparison between actual and predicted power flows.}
\label{fig:actual_predicted_scatter}
\end{figure}

The trained model achieves excellent prediction performance on the testing dataset, with an MSE of 0.001253, RMSE of 0.010080, and an $R^2$ value of 0.999997. The comparison between the actual and predicted branch power flows is illustrated in Fig.~\ref{fig:actual_predicted_scatter}, where the predicted values closely overlap the ground-truth results, indicating that the proposed neural network accurately captures the nonlinear relationship between system operating conditions and branch power flows.

\begin{figure}[t]
\centering
\includegraphics[width=0.78\textwidth]{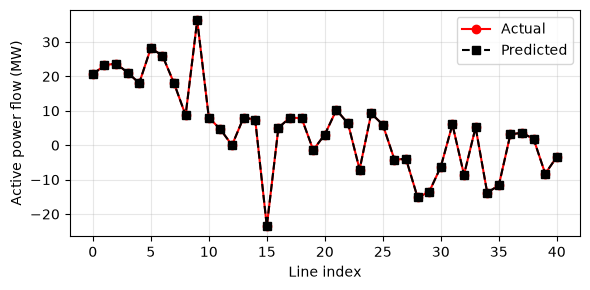}
\caption{Actual and predicted branch power flows for the first testing scenario.}
\label{fig:first_scenario}
\end{figure}

To further demonstrate the estimation accuracy, Fig.~\ref{fig:first_scenario} compares the actual and predicted power flows for all 41 transmission branches in the first testing scenario. The predicted values closely match the corresponding power flow solutions across all branches, confirming the effectiveness of the proposed framework when complete PMU measurements are available. This baseline serves as the upper performance bound for the subsequent case studies with reduced PMU deployment.

\subsection{Impact of PMU quantity}

This section investigates the impact of PMU availability on branch power flow estimation. Five PMU deployment scenarios are considered by reducing the number of monitored buses, as summarized in Table~\ref{tab:cases}. Cases 1--3 assume that 25 of the 30 buses are monitored, whereas Cases 4 and 5 reduce the number of monitored buses to 20. Among these scenarios, Cases 1 and 4 remove PMUs from one or more generator buses, while the remaining cases remove PMUs only from load buses. These case studies enable a systematic evaluation of how both the number and the locations of PMUs influence estimation performance.

\begin{table}[t]
\caption{Summary of the five PMU placement case studies.}
\label{tab:cases}
\centering
\small
\begin{tabularx}{\textwidth}{@{}lXc@{}}
\toprule
Case study & Locations without PMUs & Total monitored buses \\
\midrule
Case 1 & Generator buses: 2, 13, 22, 23, 27 & 25/30 \\
Case 2 & Load buses: 3, 8, 14, 19, 20 & 25/30 \\
Case 3 & Load buses: 3, 5, 8, 11, 25 & 25/30 \\
Case 4 & Load buses: 3, 4, 8, 10, 12, 14, 17, 20, 24; generator bus: 27 & 20/30 \\
Case 5 & Load buses: 2, 5, 7, 9, 11, 12, 17, 21, 25, 28 & 20/30 \\
\bottomrule
\end{tabularx}
\end{table}

The estimation performance for each PMU deployment scenario is summarized in Table~\ref{tab:performance}. Among the five cases, Case 3 achieves the best performance, with an RMSE of 1.889 and an $R^2$ value of 0.8802, indicating that the selected PMU locations provide highly informative measurements for reconstructing system-wide branch power flows. In contrast, Case 1 produces the largest prediction error (RMSE = 4.490, $R^2=0.3232$), even though it uses the same number of PMUs as Cases 2 and 3. This result suggests that removing PMUs from generator buses significantly degrades estimation accuracy because generator measurements contain critical information about the operating state of the network. Furthermore, Case 5, which uses only 20 PMUs, outperforms Case 1 despite having fewer measurements, demonstrating that PMU placement is more important than the number of installed PMUs. These results highlight that simply increasing the number of PMUs does not guarantee better estimation performance; instead, selecting electrically informative measurement locations is essential. The following section investigates the importance of individual buses to identify the most effective PMU locations for power flow estimation.

\begin{table}[t]
\caption{Performance comparison of the five PMU placement case studies.}
\label{tab:performance}
\centering
\begin{tabular}{@{}lccc@{}}
\toprule
Case study & Monitored buses & RMSE & $R^2$ \\
\midrule
Case 1 & 25/30 & 4.490 & 0.3232 \\
Case 2 & 25/30 & 2.066 & 0.8567 \\
Case 3 & 25/30 & 1.889 & 0.8802 \\
Case 4 & 20/30 & 2.898 & 0.7180 \\
Case 5 & 20/30 & 2.256 & 0.8290 \\
\bottomrule
\end{tabular}
\end{table}

\subsection{Impact of PMU locations}

The locations of PMUs play a crucial role in the accuracy of the proposed neural-network-based power flow estimator. To quantify the contribution of each measurement location, a feature importance analysis was performed using the RMSE increase method. In this analysis, the measurements associated with each bus were removed individually, and the resulting increase in RMSE was used to measure the importance of that bus. A larger RMSE increase indicates that the corresponding bus provides more informative measurements for reconstructing system-wide branch power flows. As summarized in Table~\ref{tab:importance}, buses 13, 2, 8, 27, and 22 are identified as the most influential locations, producing the largest increases in RMSE when excluded. These buses contain critical information about the network operating conditions and therefore have the greatest impact on estimation accuracy.

\begin{table}[t]
\caption{Top 10 most important buses based on RMSE increase analysis.}
\label{tab:importance}
\centering
\begin{tabular}{@{}cccc@{}}
\toprule
Rank & Bus & Number of features & RMSE increase \\
\midrule
1 & 13 & 1 & 3.694420 \\
2 & 2  & 3 & 3.596552 \\
3 & 8  & 2 & 2.582403 \\
4 & 27 & 1 & 2.514652 \\
5 & 22 & 1 & 1.956464 \\
6 & 7  & 2 & 1.791676 \\
7 & 23 & 3 & 1.782317 \\
8 & 21 & 2 & 1.586631 \\
9 & 30 & 2 & 1.104631 \\
10 & 19 & 2 & 0.910393 \\
\bottomrule
\end{tabular}
\end{table}

The feature importance analysis also explains the performance differences observed in Section~3.2. Although Cases 1, 2, and 3 all employ 25 PMUs, Case 1 yields the poorest prediction performance because several highly influential buses (2, 13, 22, 23, and 27) are not monitored. In contrast, Case 3 retains measurements from more informative buses and achieves the highest estimation accuracy. These results demonstrate that PMU placement is more important than the number of installed PMUs for learning-based power flow estimation. Consequently, PMUs should be installed at electrically informative buses to achieve an effective balance between deployment cost and estimation accuracy.

\section{Conclusions}

This study proposed a learning-based framework for estimating system-wide branch power flows using limited PMU measurements. The proposed neural network was validated on the IEEE 30-bus test system with 50,000 operating scenarios and achieved high estimation accuracy under full PMU deployment (RMSE = 0.010, $R^2=0.999997$). The results further demonstrated that estimation accuracy depends not only on the number of installed PMUs but also on their locations. Feature importance analysis identified the most informative buses for power flow estimation and showed that strategically placing PMUs at electrically important buses provides higher accuracy than simply increasing the number of PMUs. These findings provide practical guidance for designing cost-effective PMU deployment strategies that balance installation cost and estimation performance.

\section*{Acknowledgements}

This work was supported in part by the U.S. National Science Foundation (NSF) under Award 2552448.

\bibliographystyle{unsrt}
\bibliography{references}

\end{document}